\documentclass[pra,onecolumn,aps]{revtex4}

\usepackage{amssymb}
\usepackage{amsmath}
\usepackage[pdftex]{graphicx}

\begin{document}

\title{Micropropagation of three genotypes of Indian mustard [\textit{Brassica 
juncea} (L.) Czern.] using seedling-derived transverse thin cell layer (tTCL) 
explants.}

\author{ Michel AOUN}\email{ michel.aoun@univ-brest.fr}
\author{Gilbert CHARLES}
\author{Annick HOURMANT}
\affiliation{Laboratoire de Biotechnologie et Physiologie V\'eg\'etales, 
Universit\'e de Bretagne Occidentale (Brest - France).}

\begin{abstract}

Micropropagation of three genotypes of Indian mustard [\textit{Brassica 
juncea} (L.) Czern.] using 7-days old seedling-derived transverse thin cell 
layer (tTCL) explants was accomplished.   \\
The genotype, explant source and 
addition of silver nitrate to the medium significantly influenced shoot bud 
induction. MS medium with 26.6 $\mu$M of 6-Benzylaminopurin (BAP) and 3.22 
$\mu$M of 1-naphtaleneacetic acid (NAA) was identical (in the case of cotyledon 
tTCLs whatever the organ) and superior for the induction of buds (in the cases 
of petiole tTCL explants of genotypes 1 and 2 and hypocotyl tTCL explants of 
genotypes 1 and 3) than 53.3 $\mu$M of BAP and 3.22 $\mu$M of NAA. However, 
53.3 $\mu$M of BAP was superior for the induction of buds than 26.6 $\mu$M in 
the presence of the same concentration of NAA for petiole tTCL explants of 
genotype 3 and hypocotyl tTCL explants of genotype 2.  \\ 
The addition of silver 
nitrate significantly enhanced the rate of shoot induction in all genotypes. 
Cotyledon-derived tTCL explants exhibited the highest shoot bud induction 
potential and was followed by petiole- and hypocotyl-derived ones. Addition of 
10 $\mu$M of silver nitrate to BAP and NAA supplemented medium induced higher 
frequency shoot bud induction (up to 100 \%) with the highest means of 4.45 
shoots per cotyledon-derived tTCL explants obtained with the genotype 2. Shoot 
regenerated were rooted on MS basal medium without PGRs which induced 99 \% of 
roots per shoot. The plantlets established in greenhouse conditions with 99 \% 
survival, flowered normally and set seeds.  \\

\end{abstract}

\maketitle 


\textbf{Abbreviations} BAP 6-Benzylaminopurin; NAA 1-naphtaleneacetic acid; 
AVG (aminoethoxyvinylglycin), tTCLs transverse Thin Cell Layers.

\section{Introduction}

\textit{Brassica} oilseed crops including, \textit{Brassica napus}, \textit{B. 
rapa} and \textit{B. juncea,} cover approximately 11 million hectares of the 
world's agricultural land and provide over 8\% of the major oil grown under a 
variety of climatic conditions (Downey, 1990). The annual production of 
\textit{Brassica juncea} in India alone reaches 5 million tons (FAO, 2003). 
\textit{Brassica juncea} is widely grown as a vegetable but also for its uses 
in the food industry, in condiments, vegetable oils, hair industry, in 
lubricants and, in some countries, as a subtitute for olive oil. Seed residues 
are used as cattle feed and in fertilizers (Reed, 1976).  \\

For the sake of crop improvement, \textit{Brassica }species have been 
subjected to many genetic manipulations with \textit{Agrobacterium} species 
(Barfield and Pua, 1991; Jonoubi \textit{et al.} 2005). However, it is better 
to obtain a high rate of shoot regeneration in these species for \textit{in 
vitro} selection and routine genetic manipulation. Shoot regeneration 
in\textit{ Brassica juncea} has been reported from mesophyll protoplasts 
(Chatterjee \textit{et al.} 1985), and traditional explants such as hypocotyls 
(Sethi \textit{et al. }1990), petioles (Pua and Chi, 1993), leaf discs and 
peduncle explants (Eapen and George, 1996, 1997), cotyledon or leaf segments 
(Guo \textit{et} \textit{al.} 2005) and microspores (Prem \textit{et al.} 
2005). However, to the best of our knowledge there is no report on shoot 
regeneration of \textit{B. juncea} using explant-derived from young plants.  \\

Shoot regeneration is affected by many factors like genotype, explant source, 
medium, growth hormones and physical conditions (Jain \textit{et al.} 1988; 
Zhang \textit{et} \textit{al.} 1998; Tang \textit{et} \textit{al.} 2003; Guo 
\textit{et} \textit{al.} 2005).  \\

\textit{In vitro} culture of longitudinal thin cell layer explants, developed 
by Tran Thanh Van (1973), enabled regeneration in many species such as 
\textit{Brassica napus} (Klimaszewska and Keller, 1985), \textit{Beta vulgaris} 
(Detrez \textit{et} \textit{al.} 1988), \textit{Lupinus mutabilis} and 
\textit{L. albus} (Mullin and Bellio-Spataru, 2000).  \\

The use of thin transverse sections has been successfully applied to 
\textit{Panax ginseng} (Ahn \textit{et} \textit{al.} 1996), \textit{Digitaria 
sanguinalis} (Bui \textit{et} \textit{al.} 1997), \textit{Lilium longiflorum} 
(Nhut \textit{et} \textit{al.} 2002), \textit{Spinacia oleracea} (L\'eguillon 
\textit{et} \textit{al.} 2003) or \textit{Brassica napus} (Ben Ghnaya 
\textit{et al.} 2008). Such technique with high reactivity and sensibility, 
seems to have a potential for intensive caulogenesis and regeneration of 
\textit{Brassica juncea} plants.  \\

In this study we investigate the effect of silver nitrate on the efficiency of 
shoot regeneration from cotyledon-, petiole- and hypocotyl-derived tTCL 
explants of three genotypes of\textit{ Brassica  juncea }(L.) Czern.  \\

\section{Materials and Methods}

\textbf{ Plant material} \\

Seeds of \textit{Brassica juncea} I39/1 (genotype 1), AB79/1 (genotype 2) and 
J99 (genotype 3), provided by `Ecole Nationale d'Enseignement Sup\'erieur 
Agronomique (ENESA)' in Dijon (France) were used to evaluate shoot regeneration 
in presence or the absence of AgNO$_3$. These genotypes are pure spring lines, 
genetically fixed and were obtained by autofertilization. The choice of 
concentrations of BAP and NAA is based on preliminary results in our laboratory 
that give relatively the best rate of shoot regeneration.  \\

\textbf{ Culture conditions and plant regeneration} \\

Seeds of \textit{Brassica juncea} were decontaminated in 70\% ethanol for 30 
sec, followed by immersion in calcium hypochlorite (5\%, w/v) added with two 
drops of Tween-20 for 10 min. The seeds were rinsed twice for 5 min with 
sterile water upon sterilization. For germination, one seed was placed per 
culture tube containing 0.6 \% agar-solidified MS basal medium (Murashige and 
Skoog, 1962) and 2 \% sucrose at 5.8 pH.   \\

For the regeneration studies, tTCLs (400-500 $\mu$m thick) were excised from 
hypocotyls, petioles and cotyledons of 7 day-old plantlets. They were cultured 
in Petri dishes (15 tTCLs / dish) on MS medium supplemented with agar 0.6 \% 
(w/v), sucrose 2\% (w/v), NAA (3.22 $\mu$M), BAP (26.6-53.3 $\mu$M) and 
AgNO$_3$ (0-20 $\mu$M). All experiments were conducted in a culture chamber 
with a 12h photoperiod (60 $\mu$mol photon.m$^{-2}$.s$^{-1}$) provided by cool 
white fluorescent lamps with a 22/20${}^\circ$C thermoperiod (light/dark). The 
pH was adjusted to 5.8 with NaOH before autoclaving at 121${}^\circ$C for 20 
min.  \\

After three or four weeks, regenerated shoots were transferred to 0.7 \% 
agar-solidified MS medium without any growth hormones. After two weeks, rooted 
plants were hardened under greenhouse conditions until flowering. \\

\textbf{ Data analysis} \\

The number of explants that showed bud formation, as well as the number of 
buds per tTCL were counted and the frequency of bud formation was calculated.  \\ 

There were 15 tTCLs per Petri dish and four plates per organ and treatment. 
Each experiment was repeated 3 times with 3 independent runs.  \\

For data analysis, one way ANOVA and LSD (Least Significance difference) test 
were performed using the Statgraphics computer program at P $<$ 0.05 or 0.01 
(Statgraphics Plus version 5.1).  \\

\section{Results}

Results indicate that shoot regeneration ability is strongly influenced by 
genotype and tTCL explants between and within the \textit{Brassica juncea} 
genotypes (Table 1). The first shoots were observed after ten days on tTCLs, 
whatever the genotype, and reached their largest number of occurrence at 
day-27. Callogenesis was the first observed phenomenon, followed by 
rhizogenesis and caulogenesis respectively. Whatever the origin of tTCL 
explants, high rates of callogenesis and rhizogenesis were observed in calli 
culture (up to 99 \%) (Figure 2 A, B and C). The caulogenesis frequency ranged 
from 0 to 1 \% for gentoype 3; 6.67 to 30.47 \% for genotype 2 and 0 to 78.67 
\% for genotype 1. Furthermore, the origine of tTCL explants also affected 
shoot regeneration; cotyledon tTCLs best responded compared to petiole and 
hypocotyl tTCLs respectively and that whatever the genotype (Table 1). The best 
number of buds per tTCL explant was obtained with hypocotyl tTCLs of genotype 2 
cultivated on MS medium added with 26.6 $\mu$M BAP, 3.22 $\mu$M NAA and sucrose 
2\% (w/v) (Table 2). \\

The presence of AgNO$_3$ in the culture medium showed a significantly 
beneficial effect on shoot regeneration. It influenced both the frequency of 
shoot regeneration (Table 1) and the number of buds per tTCL explant (Table 2). 
In MS medium added with 26.6 $\mu$M BAP and 10 $\mu$M AgNO$_3$, \textit{B. 
juncea} AB79/1 (genotype 2) does not show any significant difference between 
the frequencies of shoot formation between organs,which reach up to 100 \% 
(Table 1). However, in the same conditions, the highest number of shoots per 
explant was obtained with hypocotyl tTCLs  (Table 2).  \\

In MS medium added with 26.6 $\mu$M BAP and 10 $\mu$M AgNO$_3$, \textit{B. 
juncea} I39/1 (genotype 1) and \textit{B. juncea} J99 (genotype 3), cotyledon 
tTCLs were more sensitive for shoot regeneration than those of petiole and 
hypocotyl respectively (Table 1). Furthermore, their number of buds per tTCL 
explants rose from 1 to approximatively 2 when 10 $\mu$M AgNO$_3$ were added to 
culture medium. The addition of 53.3 $\mu$M BAP and 10 $\mu$M AgNO$_3$ to 
culture medium seems to be more sensitive for given a highest frequency of 
shoot regeneration for petiole tTCL explants (19.07 \%) than 26.6 $\mu$M BAP 
(1.13 \%) for genotype 3.  \\

Moreover, 1 $\mu$M AgNO$_3$ increases significantly caulogenesis for genotypes 
1 and 2, but not for genotype 3. This more recalcitrant genotype required 10 
$\mu$M AgNO$_3$ to have a significant increase on bud formation (Table 1). 
Concentrations of AgNO$_3$ higher than 10 $\mu$M (15 and 20 $\mu$M) do not show 
any significant difference compared to 10 $\mu$M whatever the genotype tested 
(data not shown).  \\

In addition, in the presence of silver nitrate, calli were compact and had 
dark green color (Figure 1B). However, in its absence, calli were soft and had 
a pale green color whatever the concentration of growth regulators and the 
origin of tTCL explants (Figure 1A).  \\

Finally, after 27 days of tTCL culture, shoots excised and transferred in the 
test tubes on 0.7 \% agar-solidified MS medium without PGRs exhibited rooting 
and rapid development (Figure 2D). After 2 weeks, plantlets were transferred to 
grow in pots under greenhouse conditions. The phenotype of regenerant plants 
were similar to plant control grown from seeds. They flowered normally 6 weeks 
after transfer in pots and set seeds (Figure 2E).   \\

\section{Discussion and Conclusion}

Thin cell layer technology was known efficient for the propagation of various 
plant species. This study was conducted to achieve a high rate of regeneration 
in \textit{B. juncea} L. Czern. From tTCL explants in presence or in the 
absence of silver nitrate. This technique combined to AgNO$_3$ promoted rapid 
and high frequency of shoot regeneration with shoot buds developing within 10 
days for all \textit{B. juncea} genotypes.  \\

These results compare favourably with recent studies of shoot regeneration of 
\textit{B. napus} L. from traditional explants (Tang \textit{et al.} 2003; 
Akasaka-Kennedy \textit{et al.} 2005), longitudinal thin cell layers 
(Klimaszewska and Keller, 1985) and transverse thin cell layers (Ben Ghnaya 
\textit{et al.} 2008). In our study, tTCL explants were excised transversally 
from 7-day old axenic plants.  \\

In our experiments, all factors evaluated (genotype, explant, BAP and 
AgNO$_3$) influenced shoot regenration. Shoot regeneration ability is strongly 
influenced by genotype as proved earlier in \textit{B. napus} L. (Ono 
\textit{et} \textit{al.} 1994; Akasaka-Kennedy \textit{et} \textit{al.} 2005; 
Ben Ghnaya \textit{et al.} 2008) and \textit{B. campestris} L. ssp. Pekinensis 
(Zhang \textit{et} \textit{al.} 1998). \textit{B. juncea} AB79/1 (genotype 2) 
and \textit{B. juncea }I39/1 (genotype 1) showed a greater capacity to produce 
shoots on the MS medium containing 3.22 $\mu$M NAA, 26.6 or 53.3 $\mu$M BAP and 
sucrose 2 \% (w/v) than \textit{B. juncea} J99 (genotype 3). Furthemore, we 
showed an explant effect on shoot regeneration process. For all genotypes, 
cotyledon tTCLs best responded than petiole and hypocotyl tTCLs and exhibited 
the highest shoot regeneration rate (Table 1). Tang et al. (2003) and more 
recently Ben Ghnaya \textit{et al.} (2008) showed that PGR content affected 
significantly the regeneration process as observed from tTCL explants in our 
study. Whatever the genotype and organ, 26.6 $\mu$M BAP seems to be more 
sensitive on shoot regeneration frequency than 53.3 $\mu$M.  \\

Moreover, silver nitrate showed to be significantly beneficial to the shoot 
regeneration process for all genotypes of \textit{Brassica juncea}, even for 
the more recalcitrant one (genotype 3). Indeed, 10 $\mu$M AgNO$_3$ was able to 
induce the response of genotype 3 but concentrations up to more than 10 $\mu$M 
(15 and 20 $\mu$M, data not shown) didn't increase the frequencies of 
caulogenesis (Table 1).The positive effect of silver nitrate on shoot 
regenration process, was shown in previous studies with traditional explants 
such as cotyledons of \textit{Brassica rapa} ssp. Oleifera (Burnett \textit{et} 
\textit{al.} 1998), \textit{Brassica campestris} ssp. Pekinensis (Chi 
\textit{et} \textit{al.} 1991; Zhang \textit{et al.} 1998), hypocotyls of 
\textit{Brassica juncea} (Pua and Chi, 1993), peduncle and leaf segments of 
\textit{Brassica napus }(Eapen and George, 1997; Akasaka-Kennedy \textit{et} 
\textit{al.} 2005). Furthermore, the presence of silver nitrate, especially 10 
$\mu$M seems to be sensitive significantly in the enhancement of the number of 
buds per tTCL explants, whatever the genotype and the organ (Table 2).  \\

In the present study, an original and efficient regeneration system from 
cotyledon, petiole and hypocotyl tTCLs has been developed in \textit{Brassica 
juncea}. It produces good results with tTCL explants of all genotypes. Despite 
a smaller surface and a larger number of wounded cell, shoot regeneration is 
obtained typically ten days after the tTCL explant initiation culture. This 
swift response, in agreement with the observation described previously by Tran 
Thanh Van (1973), due to the combined process of cell dedifferentiation and 
reprogramming. Typically, two months later, we observed the normal flowering of 
regenerant plantlets. Nonetheless, in this system, a single subculture step 
preceding the regenerated plant transfer into culture tube is required and no 
subsequent phenotypic alterations were observed in these plants at all.  \\

For further improvements, other factors could be taken into account, such as 
the concentration of sugar used (Ben Ghnaya \textit{et al.} 2008), the use of 
other inhibitors that block ethylene synthesis (e.g. AVG, nickel and cobalt) or 
action (silver thiosulfate) (Chraibi \textit{et al.} 1991; Burnett \textit{et 
al.} 1994), but also tTCL explant thickness or position along the organ (Nhut 
\textit{et al.} 2001). \\

Our tTCL model could be used as a tool for fundamental regeneration studies 
and for crop improvement using \textit{Agrobacterium} transformation of 
\textit{Brassica juncea} cultivars as well as other \textit{Brassica }species. 
Furthermore this system may will be used for \textit{in vitro} selection, in 
presence of many metals, of plants which may will be used in different 
phytoremediation processes. \\

\vspace{1cm} 

\textbf{Acknowledgements} \\

We thank Dr T. Guinet (ENESA) for providing seeds of spring lines of 
\textit{Brassica juncea}, Pr L. Quiniou [Institut Europ\'een de la Mer (IUEM)] 
and Dr P. Rey [Ecole Sup\'erieure de Microbiologie Industrielle et de 
S\'ecurit\'e Alimentaire de Brest (ESMISAB)] for statistical analysis.

\vspace{1cm}

 \textbf{References} \\

Ahn, I.O., Bui, V.L., Gendy, C. , Tran Thanh Van, K., 1996. Direct somatic 
embryogenesis through thin cell layer culture in \textit{Panax ginseng}. 
\textit{Plant Cell Tiss. Org. Cult}. 45, 237-243.

Akasada-Kennedy, Y., Yoshida, H. , Takahata, Y., 2005. Efficient plant 
regeneration from leaves of rapeseed (\textit{Brassica napus} L.) The influence 
of AgNO$_3$ and genotype. \textit{Plant Cell Rep.} 24, 649-654.

Barfield, D.G., Pua, E.C., 1991. Gene transfer in plants of \textit{Brassica 
juncea} using \textit{Agrobacterium tumefaciens}-mediated transformation. 
\textit{Plant Cell Rep. }10, 308-314.

Ben Ghnaya, A., Charles, G., Branchard, M., 2008. Rapid shoot regeneration 
from thin cell layer explants excised from petioles and hypocotyls in four 
cultivars of \textit{Brassica napus} L. \textit{Plant Cell Tiss. Organ Cult.} 
92, 25-30.

Bui, V.L., Do My, N.T., Gendy, C., Vidal, J. , Tran Thanh Van, K., 1997. 
Somatic embryogenesis on thin cell layers of a C4 species, \textit{Digitaria 
sanguinalis} (L.) Scop. \textit{Plant Cell} \textit{Tiss. Org. Cult.} 49, 
201-208.

Burnett, L., Arnoldo, M., Yarrow, S., Huang, B., 1994. Enhancement of shoot 
regeneration from cotyledon explants of \textit{Brassica rapa} ssp. Oleifera 
through pretreatment with auxine and cytokinin and use of ethylene inhibitors. 
\textit{Plant Cell Tiss. Org. Cult.} 35, 253-258.

Chatterjee, G., Sikdar, S.R., Das, S., Sen, S.K., 1985. Regeneration of 
plantlets from mesophyll protoplasts of \textit{Brassica juncea} (L.) Czern. 
\textit{Plant Cell Rep.} 4, 245-247.

Chi, G.-L., Pua, E.-C., Goh, C.-J., 1991. Role of ethylene on \textit{de novo} 
shoot regeneration cotyledonary explants of \textit{Brassica campestris} 
pekinensis (Lour) Olsson \textit{in vitro}. \textit{Plant Physiol.} 96, 178-183.

Chraibi, K.M.B., Latche, A., Roustan, J., Fallot, J., 1991. Stimulation of 
shoot regeneration from cotyledons of \textit{Helianthus annuus} by the 
ethylene inhibitors, silver and cobalt. \textit{Plant Cell Rep.} 10, 204-207.

Detrez, C., Tetu, T., Sangwan, R.S., Sangwan-Norreel, B.S., 1988. Direct 
organogenesis from petiole and thin cell layer explants in sugar beet cultured 
\textit{in vitro}. \textit{J. Exp. Bot.} 39, 917-926.

Downey, R.K., 1990. \textit{Brassica} oilseed breeding-achievements and 
opportunities. \textit{Plant Breed.} 60, 1165-1170.

Eapen, S., George, L., 1996. Enhancement in shoot regeneration from leaf discs 
of \textit{Brassica juncea} L. Czern. and Coss. By silver nitrate and silver 
thiosulfate. \textit{Physiol. Mol. Biol. Plants} 2, 83-86.

Eapen, S., George, L., 1997. Plant regeneration from peduncle segments of oil 
seed \textit{Brassica} species Influence of silver nitrate and silver 
thiosulfate. \textit{Plant Cell Tiss. Org. Cult.} 51, 229-232.

Food and Agriculture Organization (FAO) 2003. \textit{FAO bulletin of 
statistics}, Vol. 4 No 2.

Guo, D.-P., Zhu, Z.-J., Hu, X.-X., Zheng, S.-J., 2005 Effects of cytokinins on 
shoot regeneration from cotyledon and leaf segment of stem mustard 
\textit{Brassica juncea} var. Tsatsai. \textit{Plant Cell Tiss.Org.Cult.} 83, 
123-127.

Jain, R.K., Chowdhury, J.B., Sharma, D.R., Friedt, W., 1988. Genotypic and 
media effects on plant regeneration from cotyledon explant cultures of some 
\textit{Brassica} species. \textit{Plant Cell Tissue Org. Cult.} 14, 197-206.

Jonoubi, P., Mousavi, A., Majd, A., Salmanian, A.H., Jalali Javaran, M., Dane 
Shian, J., 2005. Efficient regeneration of \textit{Brassica napus} L. 
hypocotyls and genetic transformation by \textit{Agrobacterium tumefaciens. 
Biol Plant.} 49, 175-180.

Klimaszewska, K., Keller, W.A., 1985. High frequency plant regeneration from 
thin layer explants of \textit{Brassica napus}. \textit{Plant Cell Tiss. Organ 
Cult.} 4, 183-197.

L\'eguillon, S., Charles, G., Branchard, M., 2003. Plant regeneration from 
thin cell layers in \textit{Spinacia oleracea}. \textit{Plant Cell Tiss. Org. 
Cult.} 74, 257-265. 

Mullin, M., Bellio-Spataru, A., 2000. Organogenesis from hypocotyl thin cell 
layers of \textit{Lupinus mutabilis} and \textit{L. albus}. \textit{Plant 
Growth Reg}. 30, 177-183.

Murashige, T., Skoog, F., 1962. Revised medium for rapid growth and bioassay 
with tobacco tissue cultures. \textit{Physiol. Plant.} 15, 473-497.

Nhut, D.T., Bui, V.L., Fukai, S., Tanaka, M., Tran Thanh Van, K., 2001. 
Effects of activated charcoal, explant size, explant position and sucrose 
concentration on plant and shoot regeneration of \textit{Lilium longiflorum} 
via young stem culture. \textit{Plant Growth Regul.} 33, 59-65.

Nhut, D.T., Bui, V.L., Tran Thanh Van, K., 2002. Somatic embryogenesis through 
pseudo-bulblet transverse thin cell layer of \textit{Lilium longiflorum}. 
\textit{Plant Growth Reg.} 37, 193-198.

Ono, Y., Takahata, Y., Kaizuma, N., 1994. Effect of genotype on shoot 
regeneration from cotyledonary explants of rapeseed (\textit{Brassica napus} 
L.). \textit{Plant Cell. Rep.} 14, 13-17.

Prem, D., Gupta, K., Agnihotri, A., 2005. Effect of various exogenous and 
endogenous factors on microspore embryogenesis in Indian mustard 
[\textit{Brassica juncea} (L.) Czern. And Coss.]. \textit{In Vitro Cell Dev. 
Biol. }-\textit{ Plant} 41, 266-273.

Pua, E.-C., Chi, G.-L., 1993. \textit{De novo} shoot morphogenesis and plant 
growth of mustard (\textit{Brassica juncea}) \textit{in vitro} in relation to 
ethylene. \textit{Physiol. Plant.} 88, 467-474.

Reed, C.F., 1976. Information summaries on 1000 economic plants, United States 
Department of Agriculture (USDA).

Sethi, U., Basu, A., Guha-Mukherjee, S., 1990. Control of cell proliferation 
and differentiation by modulators of ethylene biosynthesis and action in 
\textit{Brassica} hypocotyl explants. \textit{Plant Sci. }69, 225-229.

Tang, G.X., Zhou, W.J., Li, H.Z., Mao, B.Z., He, Z.H., Yoneyama, K., 2003. 
Medium, explant and genotype factors influencing shoot regeneration in oilseed 
\textit{Brassica} spp. \textit{J. Agron. Crop Sci.} 189, 351-358.

Tran Thanh Van, M., 1973. \textit{In vitro} control of \textit{de novo} 
flower, bud, root, and callus differentiation from excised epidermal tissues. 
\textit{Nature} (Lond.) 246, 44-45.

Zhang, F.L., Takahata, Y., Xu, J.B., 1998. Medium and genotype factors 
influencing shoot regeneration from cotyledonary explants of Chinese cabbage 
(\textit{Brassica campestris} L. ssp. Pekinensis). \textit{Plant Cell Rep.} 17, 
780-786.

\begin{figure}[htbp]
\begin{center}
\scalebox{0.8}{\includegraphics*{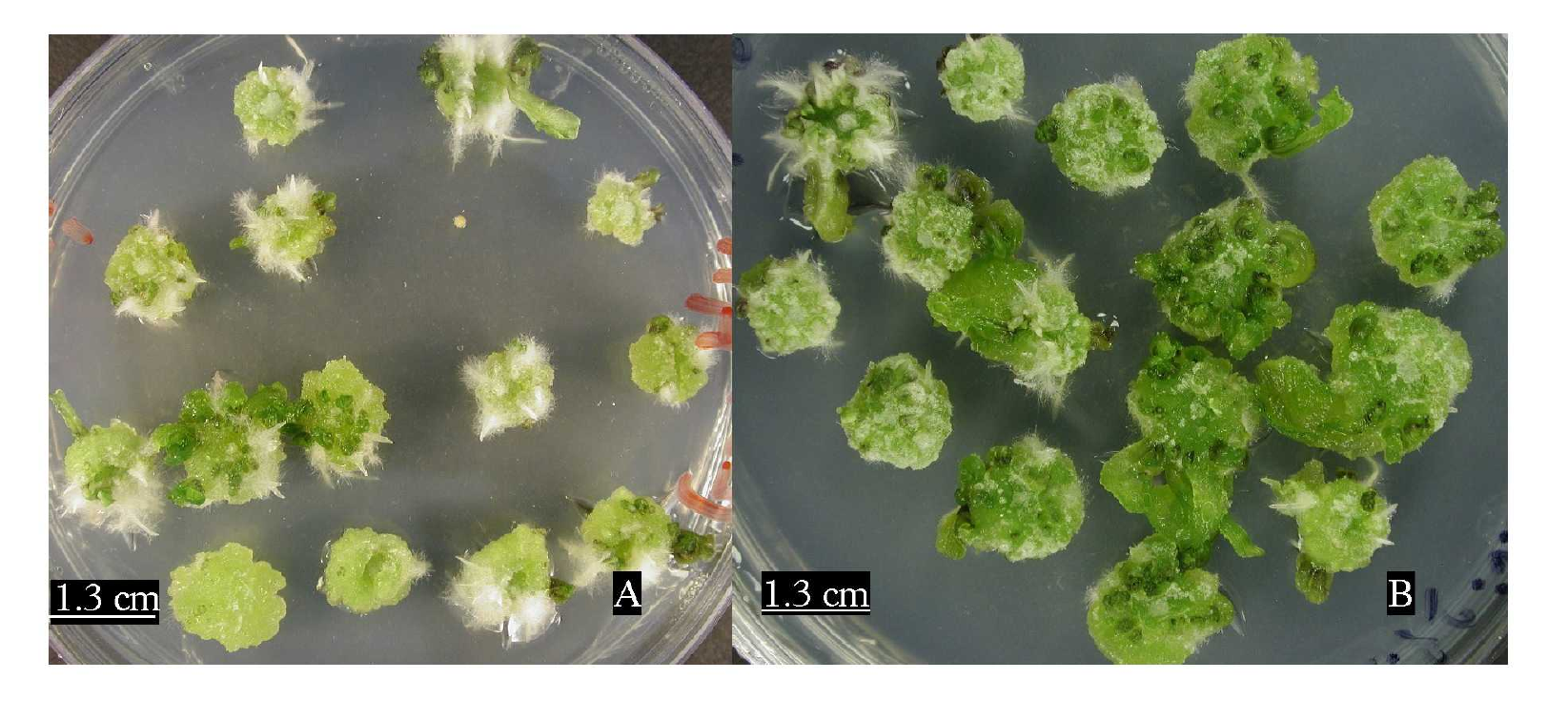}}
\end{center}
\caption{Morphology of calli obtained from hypocotyl tTCL explants of 
\textit{Brassica juncea} AB79/1 in absence (A) or in presence (B) of AgNO$_3$.} 
\label{fig1}
\end{figure}

\begin{figure}[htbp]
\begin{center}
\scalebox{0.8}{\includegraphics*{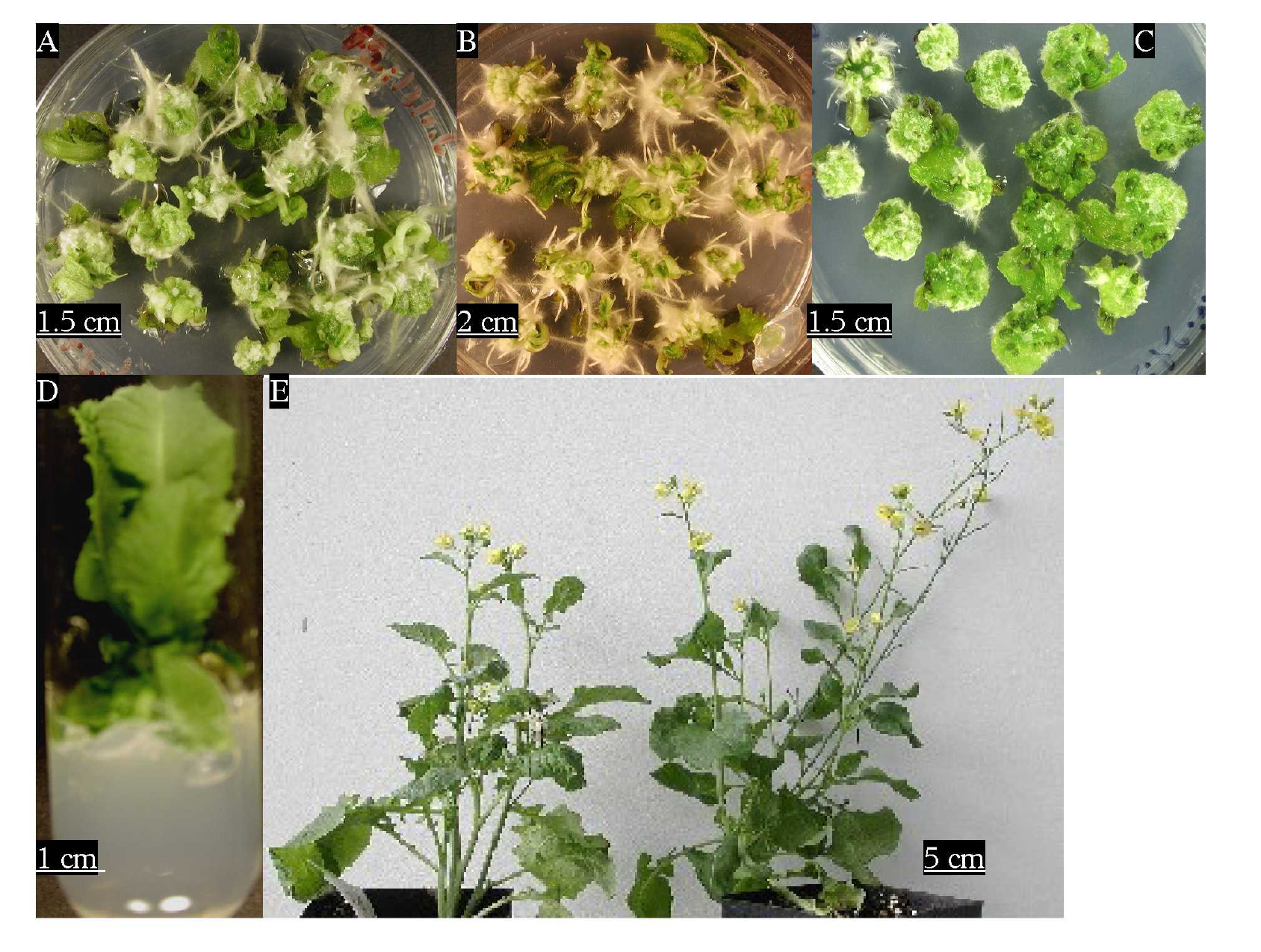}}
\end{center}
\caption{Shoot regeneration from tTCL explants of \textit{Brassica juncea} 
AB79/1. (A): cotyledon tTCLs; (B): petiole tTCLs; (C): hypocotyl tTCLs; (D): 
plantlet obtained from excised tTCL grown 2 weeks in a test tube and (E): 
regenerated plants grown and flowered in the greenhouse.}
 \label{fig2}
\end{figure}

\newpage

\begin{table}[htbp]
\begin{center}
\caption{Effect of genotype, explant, BAP and AgNO$_{3}$ on \textit{in vitro} 
organogenesis 
from tTCLs of cotyledons, petioles and hypocotyls. The percentage of buds 
were recorded 27 days after the tTCL initiation culture.} 
\label{tab1}
\begin{tabular}{ c c c c c c}
\hline   
Genotypes & AgNO$_{3 }$ & BAP & \underline {Frequency of of regenerating tTCL 
({\%})}  \\
    & & Cotyledons & Petioles & Hypocotyls  \\

\hline
1 &  0 & 26.6 & 78.67$^{c }$ & 23.34$^{fg }$ & 0.00$^{k}$   \\

&1 &&93.34$^{a }$& 68.93$^{d }$& 11.13$^{gh}$  \\

& 5 & & 96.67$^{a }$& 75.60$^{c}$&  13.34$^{gh}$  \\

& 10 && 97.34$^{a }$& 77.80$^{c }$& 18.67$^{g}$  \\

\\

& 0 & 53.3 & 72.20$^{c }$& 8.00$^{gh}$&  0.00$^{k}$  \\

& 1 && 86.67$^{b }$& 32.27$^{f }$& 7.80$^{h}$  \\

& 5 & & 93.34$^{a }$& 44.00$^{e }$& 7.80$^{h}$  \\

& 10 & &94.47$^{a }$& 46.67$^{e }$& 11.14$^{gh}$  \\

\hline

2 & 0 & 26.6 & 30.47$^{f }$& 22.27$^{fg }$& 6.67$^{h}$   \\

& 1& & 93.34$^{a }$& 87.88$^{b }$& 65.56$^{d}$  \\

& 5 & &96.67$^{a }$& 92.27$^{a }$& 87.80$^{b}$ \\

& 10 & &\textbf{100.00}$^{a }$& 98.87$^{a }$& 92.23$^{a}$ \\

\\

& 0 & 53.3 & 26.67$^{f }$& 10.00$^{gh }$& 16.67$^{g}$ \\

& 1 && 95.34$^{a }$& 67.80$^{d }$& 37.80$^{e}$ \\

& 5 && 95.60$^{a}$ & 75.60$^{c }$& 73.34$^{c}$ \\

& 10 && 96.67$^{a }$& 84.47$^{b }$& 75.60$^{c}$ \\

\hline

3 & 0&  26.6&  0.93$^{c }$& 0.00$^{k }$& 0.00$^{k}$ \\

& 1&&  0.93$^{c }$& 0.00$^{k }$& 0.93$^{j}$ \\

& 5 && 6.67$^{h}$&  0.00$^{k }$& 9.53$^{gh}$ \\

& 10&&  15.20$^{g }$& 1.13$^{j }$& 9.53$^{gh}$ \\

\\

& 0&  53.3 & 0.93$^{j }$& 0.93$^{j }$& 0.00$^{k}$ \\

& 1 && 0.93$^{j }$& 2.87$^{i }$& 0.00$^{k}$ \\

& 5 && 12.40$^{gh }$& 10.47$^{gh }$& 0.00$^{k}$ \\

& 10&&  13.34$^{gh }$& 19.07$^{g }$& 0.00$^{k}$ \\

\hline
\end{tabular}
\end{center}
\end{table}

The frequency of shoot regeneration were recorded 27 days after the tTCL 
initiation culture on MS basal medium supplemented with NAA (3.22 $\mu $M), 
BAP (26.6 or 53.3 $\mu $M), sucrose 2 {\%} (w/v) and AgNO$_{3}$ (0 - 10 $\mu 
$M). The results were calculated from three independent experiments, each 
with, at least, five Petri dishes with 15 tTCLs per dish. The mean values 
with different letters are significantly different at $p$ = 0.05 (LSD test).

\newpage

\begin{table}[htbp]
\begin{center}
\caption{Effect of genotype, explant, BAP and AgNO$_{3}$ on the number of 
adventitious buds per tTCL excised from different organs (cotyledons, 
petioles and hypocotyls).} 
\label{tab2}
\begin{tabular}{ c c c c c c}
\hline 
Genotypes & AgNO$_{3 }$ & BAP & \underline {Number of buds per tTCLs } \\
 & & Cotyledons & Petioles&  Hypocotyls \\

\hline

1 &  0&   26.6&   1.34$^{ef }$&  1.16$^{ef }$&  0$^{g}$ \\

&  1 &&  1.69$^{e }$&  1.75$^{e }$&  1.00$^{f}$ \\

&  5 &&  1.81$^{e }$&  1.86$^{e }$&  1.07$^{f}$ \\

&  10 &&  1.86$^{e }$&  1.90$^{e }$&  1.12$^{f}$ \\

\\

&  0 &  53.3 &  1.26$^{ef }$&  1.00$^{f }$&  0$^{g}$ \\

&  1 &&  1.51$^{e }$&  1.02$^{f }$&  1.00$^{f}$ \\

&  5 &&  1.82$^{e }$&  1.23$^{cd }$&  1.00$^{f}$ \\

&  10 &&  1.89$^{e }$&  1.30$^{cd }$&  1.14$^{f}$ \\

\hline

2&   0 &  26.6 &  1.00$^{f }$&  1.23$^{ef }$&  2.09$^{de}$ \\

&  1 &&  1.64$^{e }$&  2.60$^{d }$&  2.88$^{d}$ \\

&  5 &&  2.27$^{de }$&  3.16$^{c }$&  3.87$^{b}$ \\

&  10 &&  2.40$^{de }$&  3.30$^{c }$&  \textbf{4.45}$^{a}$ \\

\\

&  0 &  53.3 &  1.00$^{f }$&  1.12$^{f }$&  1.42$^{ef}$ \\

&  1 &&  1.14$^{f }$&  2.75$^{d }$&  1.68$^{e}$ \\

&  5 &&  2.15$^{de }$&  3.26$^{c }$&  2.67$^{d}$ \\

&  10 &&  2.22$^{de }$&  3.43$^{c }$&  2.78$^{d}$ \\

\hline

3&   0 &  26.6 &  1.00$^{f }$&  0$^{g }$&  0$^{g}$ \\

&  1 &&  1.00$^{f }$&  0$^{g }$&  1.00$^{f}$ \\

&  5 &&  1.23$^{ef }$&  0$^{g }$&  1.12$^{f}$ \\

&  10&&   1.34$^{ef }$&  1.00$^{f }$&  1.23$^{ef}$ \\

\\

&  0 &  53.3 &  1.00$^{f }$&  1.00$^{f }$&  0$^{g}$ \\

&  1 &&  1.00$^{f }$&  1.11$^{f }$&  0$^{g}$ \\

&  5 &&  1.27$^{ef }$&  1.31$^{ef }$&  0$^{g}$ \\

&  10 &&  1.41$^{ef }$&  1.37$^{ef }$&  0$^{g}$ \\

\hline
\end{tabular}
\end{center}
\end{table}

Data were recorded 4 weeks after the tTCL initiation culture on MS basal 
medium supplemented with NAA (3.22 $\mu $M), BAP (26.6 or 53.3 $\mu $M), 
sucrose 2 {\%} (w/v) and AgNO$_{3}$ (0 - 10 $\mu $M).

The results are obtained from three independent experiments, each with, at 
least, five Petri dishes and 15 tTCLs per dish. Letters indicate significant 
statistical differences at $p$ = 0.01 (One way ANOVA and LSD test).

\end{document}